\documentclass[a4paper, onecolumn, 11pt, accepted=2025-09-16]{quantumarticle}

\pdfoutput=1

\usepackage[utf8]{inputenc}
\usepackage[english]{babel}
\usepackage[LGR, T1]{fontenc}
\usepackage{amsmath}
\usepackage{float}
\usepackage{graphicx}
\usepackage{amsmath}
\usepackage{amsfonts}
\usepackage{amsthm}
\usepackage{amssymb}
\usepackage{etoolbox}
\usepackage[dvipsnames]{xcolor}
\usepackage{hyperref}
\usepackage{soul}
\usepackage{bbm}
\usepackage{commath}
\usepackage{enumerate}
\usepackage{mathtools}
\usepackage{calrsfs}
\DeclareMathAlphabet{\pazocal}{OMS}{zplm}{m}{n}
\usepackage{upgreek}
\usepackage{calc}
\usepackage{accents}
\usepackage{siunitx}
\usepackage{color}
\usepackage{textcomp}
\usepackage{bm}
\usepackage{dsfont}
\usepackage{relsize}
\usepackage{braket}
\usepackage{enumerate}   
\usepackage{mathrsfs}
\usepackage[symbol]{footmisc}
\usepackage[numbers, sort&compress]{natbib}
\usepackage{array}
\renewcommand*{\arraystretch}{1.1}
\newcommand*{\mline}[1]{%
\begingroup
    \renewcommand*{\arraystretch}{1.1}%
   \begin{tabular}[c]{@{}>{\raggedright\arraybackslash}p{2cm}@{}}#1\end{tabular}%
  \endgroup
}

\graphicspath{{fig/}} 

\newcommand{\declarebsfgreek}[2]{%
  \protected\csdef{bsf#1}{\mathord{\text{\bsfgreekfont#2}}}%
}
\newcommand{\bsfgreekfont}{\usefont{LGR}{cmss}{bx}{it}}
\declarebsfgreek{alpha}{a}
\declarebsfgreek{beta}{b}
\declarebsfgreek{gamma}{g}
\declarebsfgreek{delta}{d}
\declarebsfgreek{epsilon}{e}
\declarebsfgreek{zeta}{z}
\declarebsfgreek{eta}{h}
\declarebsfgreek{theta}{j}
\declarebsfgreek{iota}{i}
\declarebsfgreek{kappa}{k}
\declarebsfgreek{lambda}{l}
\declarebsfgreek{mu}{m}
\declarebsfgreek{nu}{n}
\declarebsfgreek{xi}{x}
\declarebsfgreek{omicron}{o}
\declarebsfgreek{pi}{p}
\declarebsfgreek{rho}{r}
\declarebsfgreek{sigma}{s}
\declarebsfgreek{tau}{t}
\declarebsfgreek{upsilon}{u}
\declarebsfgreek{phi}{f}
\declarebsfgreek{chi}{q}
\declarebsfgreek{psi}{u}
\declarebsfgreek{omega}{w}

\newcommand{\ignore}[1]{}
\newcommand{\nobibentry}[1]{{\let\nocite\ignore\bibentry{#1}}}

\newcommand{\bea}{\begin{eqnarray}}
\newcommand{\eea}{\end{eqnarray}}

\newcommand{\Hint}{\pmb{V}}
\newcommand{\B}{\pmb{B}}
\newcommand{\state}{\pmb\varrho}

\DeclareMathAlphabet{\pazocal}{OMS}{zplm}{m}{n}

\DeclareMathOperator{\tr}{tr}

\def\Xint#1{\mathchoice
   {\XXint\displaystyle\textstyle{#1}}%
   {\XXint\textstyle\scriptstyle{#1}}%
   {\XXint\scriptstyle\scriptscriptstyle{#1}}%
   {\XXint\scriptscriptstyle\scriptscriptstyle{#1}}%
   \!\int}
\def\XXint#1#2#3{{\setbox0=\hbox{$#1{#2#3}{\int}$}
     \vcenter{\hbox{$#2#3$}}\kern-.5\wd0}}

\def\dashint{\Xint-}

\usepackage{environ}
\makeatletter
\NewEnviron{quantifiedequation}[1]{
  \begin{equation}
  \expandafter\make@quantifiedequation\expandafter{\BODY}{#1}
  \end{equation}
}
\newcommand{\make@quantifiedequation}[2]{%
  \m@th 
  \sbox\z@{$\qquad\qquad\displaystyle#2$}
  \sbox\tw@{\let\label\@gobble$\displaystyle#1$}
  \ifdim\dimexpr 1em+\wd\z@+0.5\wd\tw@+2em>0.5\displaywidth
    #2\qquad#1
  \else
    \makebox[0pt][r]{%
      \makebox[\dimexpr0.5\displaywidth-0.5\wd\tw@][l]{\quad\box\z@}%
    }#1
  \fi
}
\makeatother

\renewcommand{\thefootnote}{\arabic{footnote}}

\theoremstyle{plain}

\numberwithin{obs}{section}

\newcommand{\ba}{\begin{align}}
\newcommand{\ea}{\end{align}}

\makeatletter
\DeclareFontFamily{OMX}{MnSymbolE}{}
\DeclareSymbolFont{MnLargeSymbols}{OMX}{MnSymbolE}{m}{n}
\SetSymbolFont{MnLargeSymbols}{bold}{OMX}{MnSymbolE}{b}{n}
\DeclareFontShape{OMX}{MnSymbolE}{m}{n}{
	<-6>  MnSymbolE5
	<6-7>  MnSymbolE6
	<7-8>  MnSymbolE7
	<8-9>  MnSymbolE8
	<9-10> MnSymbolE9
	<10-12> MnSymbolE10
	<12->   MnSymbolE12
}{}
\DeclareFontShape{OMX}{MnSymbolE}{b}{n}{
	<-6>  MnSymbolE-Bold5
	<6-7>  MnSymbolE-Bold6
	<7-8>  MnSymbolE-Bold7
	<8-9>  MnSymbolE-Bold8
	<9-10> MnSymbolE-Bold9
	<10-12> MnSymbolE-Bold10
	<12->   MnSymbolE-Bold12
}{}


\begin{document}

\title{Potential renormalisation, Lamb shift and mean-force Gibbs state---to shift or not to shift?}

\author{Luis A. Correa}
\affiliation{Instituto Universitario de Estudios Avanzados (IUdEA)}
\email{lacorrea@ull.edu.es}
\affiliation{Sección de Física, Facultad de Ciencias, Universidad de La Laguna, La Laguna 38203, Spain}
\affiliation{Department of Physics and Astronomy, University of Exeter, Exeter EX4 4QL, United Kingdom}

\author{Jonas Glatthard}
\affiliation{Department of Physics and Astronomy, University of Exeter, Exeter EX4 4QL, United Kingdom}
\email{J.Glatthard@exeter.ac.uk}

\begin{abstract}
Often, the microscopic interaction mechanism of an open quantum system gives rise to a `counter term' which renormalises the system Hamiltonian. Such term compensates for the distortion of the system's potential due to the finite coupling to the environment. Even if the coupling is weak, the counter term is, in general, not negligible. Similarly, weak-coupling master equations feature a number of `Lamb-shift terms' which, contrary to popular belief, cannot be neglected. Yet, the practice of vanishing both counter term and Lamb shift when dealing with master equations is almost universal; and, surprisingly, it can yield \textit{better} results. By accepting the conventional wisdom, one may approximate the dynamics more accurately and, importantly, the resulting master equation is guaranteed to equilibrate to the correct steady state in the high-temperature limit. In this paper we discuss why is this the case. Specifically, we show that, if the potential distortion is small---but non-negligible---the counter term does not influence any dissipative processes to second order in the coupling. Furthermore, we show that, for large environmental cutoff, the Lamb-shift terms approximately cancel any coherent effects due to the counter term---this renders the combination of both contributions irrelevant in practice. We thus provide precise conditions under which the open-system \textit{folklore} regarding Lamb shift and counter terms is rigorously justified.
\end{abstract}

\maketitle

\section{Introduction}\label{sec:intro}

Markovian master equations lay at the root of quantum thermodynamics \cite{alicki2018introduction}. They can be used to model open systems weakly coupled to various heat baths, and allow for the thermodynamically consistent definition of steady-state heat currents \cite{alicki1979engine,geva1992quantum}. In turn, this enables the design of quantum heat-engine and cooling cycles \cite{kosloff2014quantum,ghosh2018thermodynamic}, some of which have found experimental realisation on a wide range of platforms \cite{zou2017quantum,brantut2013thermoelectric,rossnagel2016single,von2019spin,thierschmann2015three,klatzow2019experimental}. Crucially, any approximations affecting the \textit{structure} of master equations can have a major impact on the subsequent thermodynamic analysis. Hence, every assumption made needs to be justified carefully.

Specifically, let us think of an arbitrary open system with Hamiltonian $\pmb{H}_\text{tot} = \pmb{H}_S + \pmb{H}_B + \pmb{H}_\text{int} $, where the terms stand for system, bath, and system--bath coupling, respectively. Here, $\pmb{H}_\text{int} = \zeta\,\pmb{V}$ and the dimensionless parameter $\zeta$ carries the order of magnitude of the coupling. Here, and in what follows, operators will be denoted by boldface symbols. Assuming that the bath is initially in equilibrium at inverse temperature $\beta=1/k\,T$, i.e., $\state_B(0) = e^{-\beta\,\pmb{H}_B}/\tr_B{e^{-\beta\,\pmb{H}_B}} = \pmb{\pi}_B $, the system is generally believed to evolve towards the mean-force Gibbs state $\pmb{\tau}_{MF} = \tr_B e^{-\beta\,\pmb{H}}/\tr e^{-\beta\,\pmb{H}}$ \cite{onsager1933theories,jarzynski2004nonequilibrium,trushechkin2022open,subacsi2012equilibrium}. For a harmonic bath, the classical limit of $\pmb{\tau}_{MF}$ is \cite{cerisola2022quantum,cresser2021weakandultrastrong,timofeev2022hmf}
{ \begin{equation}\label{eq:general-high-T-mean-force}
	\pmb{\tau}_{MF}\xrightarrow[\beta\ll\omega_0]{} \frac{e^{-\beta\,(\pmb{H}_S - Q\,\pmb{S}^2)}}{\tr_S{e^{-\beta\,(\pmb{H}_S - Q\,\pmb{S}^2)}}},
\end{equation}}
where we cast $ \pmb{H}_\text{int} = \zeta\,\pmb{V} = \pmb{S}\otimes\pmb{B} $ as the product of a system and a bath operator, and $Q$ is the \textit{reorganisation energy} \cite{wu2010efficient,ritschel2011efficient}.  { We shall assume that the small parameter $\zeta$ is absorbed into the bath coupling operator $\pmb{B}$}. From now on, we shall also take $\hbar=k=1$. Eq.~\eqref{eq:general-high-T-mean-force} differs from the local thermal state $\pmb{\pi}_S = e^{-\beta\,\pmb{H}_S} / \tr_S{e^{-\beta\,\pmb{H}_S}}$ due to the distortion on the system's potential from the finite coupling to the bath \cite{weiss1999}.   {In the open quantum systems literature, `reorganisation energy' is often taken as a measure of the system--bath interaction strength \cite{cheng2009} and it may be directly calculated from the spectral density (cf. Eq.~\eqref{eq:reorg-energy}). It takes its name from physical chemistry, in particular, the Marcus theory of electron transfer reactions \cite{marcus1985}, where `reorganisation energy' stands for the energy required to rearrange reactant and solvent molecular structure for the transfer to occur. This can be calculated via \emph{ab initio} methods such as molecular dynamics or density functional theory \cite{hsu2020}.} 

Whenever such potential distortion effect is not physically expected, instead of setting $\pmb{H}_S $ to the \textit{bare} Hamiltonian $\pmb{H}_S^{(0)}$ of the system in isolation, one puts $\pmb{H}_S = \pmb{H}_S^{(0)} + Q\,\pmb{S}^2$, which includes a \textit{counter term} and guarantees convergence towards the correct high-temperature steady state \cite{caldeira1983quantum}. Adding a counter term to the model \textit{ad hoc} has clear advantages. Namely, it keeps us from mistaking the mere potential renormalisation for other dissipative effects \cite{caldeira1983quantum}. Furthermore, this term ensures that the potential remains confining even at strong coupling \cite{weiss1999}. However, one must bear in mind that, when present, the counter term should arise \textit{naturally} from a microscopic derivation, instead of being introduced `by hand'. If present, it is thus an integral part of the {  {  microscopic Hamiltonian}} $\pmb{H}_S$ \cite{caldeira1983quantum}. Yet, when building master equations the counter term is systematically ignored \cite{correa2019pushing}, even when it is not negligible. 

Similarly, several terms are customarily removed from the Markovian master equations describing the dissipative dynamics of open systems weakly coupled to their environments; specifically, from the non-secular Born--Markov Redfield equation \cite{redfield1957theory,redfield1965theory}. These are the so-called \textit{Lamb-shift terms}, which contribute both to an effective renormalisation of the system as well as to purely dissipative processes \cite{ishizaki2009adequacy,cresser2017coarse,hartmann2020environmentally}. In order to justify their removal, one typically argues that the Lamb shift itself is small or that such terms do not have an impact on the dissipative dynamics. Unfortunately, this is not true in general \cite{ishizaki2009adequacy,thingna2012generalized,winczewski2021renormalization,trushechkin2021unified}. 

{ Rather paradoxically, these two manipulations may yield a \textit{more accurate} equation}. In this paper we set out to clarify why. Specifically, we show that the practice of 
\begin{enumerate}[(i)]
	\item renormalising the system Hamiltonian\footnote{$\pmb{H}_S$ includes the bare $ \pmb{H}_S^{(0)} $ by default and, if the model requires it, also a counter term.} $ \pmb{H}_S $ by \textit{subtracting} $ Q\,\pmb{S}^2 $,
	\item writing a conventional Born--Markov Redfield equation for such renormalised Hamiltonian,
	\item and removing all Lamb-shift terms from it,
\end{enumerate}
is supported by the rigorous derivation of a non-secular master equation to second order in the system--bath coupling. Provided that the secular approximation is justified, such equation may be further brought into the Davies or Gorini--Kossakowski--Lindblad--Sudarshan (GKLS) form \cite{davies1974markovian,gorini1976completely,lindblad1976generators}, which is most commonly used.

In order to build an intuition about why such artefact works, we first exploit the simplicity of an exactly solvable damped harmonic oscillator. Namely, we analytically show how, after undertaking steps (i)--(iii) above, our model is indeed driven towards the classical limit of $\pmb{\tau}_{MF}$ in Eq.~\eqref{eq:general-high-T-mean-force}. We further show that this limiting case is remarkably close to the exact mean-force Gibbs state over a broad range of parameters, which extends far beyond the high-temperature regime. Also, when the environmental cutoff is large, the transient oscillations are well captured.

We then show, in full generality, how a consistent derivation of the second-order non-secular Redfield equation can ultimately lead to the \textit{recipe} above. The only underlying assumptions, besides the usual weak-coupling and Born--Markov approximations, is that the reorganisation energy is of second order in the system--bath coupling (i.e., small, yet non-negligible), and that the high-frequency cutoff of the environment is sufficiently large compared with the relevant Bohr frequencies of the system. Hence, we provide a clean explanation for the unlikely success of the standard procedure, as well as means to decide whether or not it applies to any specific problem. 

This paper is organised as follows. In Sec.~\ref{sec:potential-renormalisation}, we introduce our model and describe the potential renormalisation effect and the corresponding counter term. Then, in Sec.~\ref{sec:born-markov}, we turn our attention to the Born--Markov quantum master equation and establish a link between reorganisation energy and Lamb shift. Next, and still arguing on the basis of our example model, we discuss the mutual cancellation of counter term and Lamb-shift terms (cf. Sec.~\ref{sec:to-shift-or-not-to-shift}). In Sec.~\ref{sec:proof} we will present our central result, which applies to completely general models; namely, that a rigorous perturbative treatment of the system--bath coupling strength supports the standard practice, provided that one works in the adiabatic regime of large environmental cutoff. We then go back to our toy model and solve it exactly (cf. Sec.~\ref{sec:exact}) so as to have a benchmark to illustrate our results in Sec.~\ref{sec:results}. Finally, in Sec.~\ref{sec:conclusions} we summarise and draw our conclusions.

\section{Potential renormalisation}\label{sec:potential-renormalisation}

Let us consider a harmonic oscillator with bare Hamiltonian
\begin{equation}
	\pmb H_S^{(0)} = \frac{1}{2} \omega_0^2\,\pmb x^2 + \frac{1}{2}\,\pmb p^2 \label{eq:system_Hamiltonian}
\end{equation}
and mass $m=1$. The system is coupled to the harmonic bath $\pmb H_B = \sum\nolimits_\mu \omega_\mu^2 m_\mu \pmb x_\mu^2/2  + \pmb p_\mu^2/(2 m_\mu)$ through\footnote{{ While the system coupling operator $\pmb{S}$ is, in principle, arbitrary, we implicitly assume---as it is most commonly done---that the coupling operator $\pmb{B}$ is linear in the quadratures of the bath. Also it is in this case that the distortion caused on the system potential by the system--bath coupling is captured by $Q$, as defined in Eq.~\eqref{eq:reorg-energy} below}.}
\begin{equation}\label{eq:sys-bath}
	\pmb{H}_\text{int} = \pmb x\otimes\sum\nolimits_\mu g_\mu \pmb x_\mu = \pmb x\otimes\B.
\end{equation}
The couplings $g_\mu$ are set by the the spectral density $J(\omega) = \pi\,\sum\nolimits_\mu g_\mu^2/(2 m_\mu \omega_\mu)\,\delta(\omega - \omega_\mu)$ which, in our case, is given by
\begin{equation}\label{eq:spectral_density_OhmicAlg}
	J(\omega) = \frac{\lambda\,\omega}{1 + (\omega/\Lambda)^2},    
\end{equation}
where the parameter $\lambda$ has appropriate dimensions and its magnitude is of order $\zeta^2$, thus capturing the system--bath coupling strength. In turn, $\Lambda$ is the high-frequency cutoff.

As already advanced, our exactly solvable system evolves towards the mean-force Gibbs state $\pmb{\tau}_{MF}$, whose high-temperature limit is
\begin{equation}\label{eq:mean-force-classical}
	\pmb{\tau}_{MF}(\beta\ll\omega_0) \simeq \frac{e^{-\beta\left[\frac12(\omega_0^2-\delta\omega^2)\,\pmb{x}^2 + \frac12\,\pmb{p}^2\right]}}{\tr_S\,e^{-\beta\left[\frac12(\omega_0^2-\delta\omega^2)\,\pmb{x}^2 + \frac12\,\pmb{p}^2\right]}}.
\end{equation}

This is the thermal state of an oscillator in a modified harmonic potential at the temperature of the bath. The frequency shift is $\delta\omega^2 = \frac{1}{\pi}\int_{-\infty}^\infty d\omega\,J(\omega)/\omega$, and $\frac12\,\delta\omega^2\,\pmb{x}^2$ is thus the reorganisation energy term. {  More generally, this will be given by
\begin{equation}\label{eq:reorg-energy}
    Q\,\pmb{S}^2 = \left(\sum\nolimits_\mu \frac{g_\mu^2}{2\,m_\mu\,\omega_\mu^2}\right)\,\pmb{S}^2 = \left(\frac{1}{2\pi}\dashint_{-\infty}^\infty d\omega\,\frac{J(\omega)}{\omega}\right)\,\pmb{S}^2,
\end{equation}
where $\pmb{S}$ is the system part of the system--bath coupling term and $\dashint$ stands for Cauchy principal value integral. Here, and it what follows, we extend the spectral density as an odd function for negative frequencies, i.e., $J(\omega)=-J(-\omega)$.} Importantly, the resulting shift $ \delta\omega^2 $ is not necessarily negligible \cite{correa2019pushing,winczewski2021renormalization}. For instance, $ \delta\omega^2 = \lambda\,\Lambda $ for the $J(\omega)$ in Eq.~\eqref{eq:spectral_density_OhmicAlg}, meaning that the shift is also $\pazocal{O}(\zeta^2)$. Hence, its removal is not justified when deriving a second-order master equation.

\section{Born--Markov master equations}\label{sec:born-markov}

\subsection{Redfield master equation}\label{sec:redfield}

Having introduced our model, let us tackle its dynamics. Assuming that the system--bath coupling is weak, we may approximate the time evolution through the ``Markovian'' Redfield quantum master equation \cite{bp}
\begin{equation}\label{eq:Redfield-general}
	\frac{d\widetilde{\pmb\varrho}_S}{dt} = - \zeta^2\int_0^\infty ds\,\tr_B\,[\widetilde{\pmb{V}}(t),\,[\widetilde{\pmb{V}}(t-s),\,\widetilde{\pmb\varrho}_S(t)\otimes\pmb{\pi}_B]],
\end{equation}
where $\state(0) = \state_S(0)\otimes\pmb{\pi}_B$. Hence, $\langle \pmb{V} \rangle = 0$ since $\langle\pmb{B}\rangle_B = \tr_B[\pmb{B}\,\pmb{\pi}_B] = 0$ and the resulting equation is homogeneous\footnote{Whenever $\langle \pmb{B} \rangle_B \neq 0$ one may redefine the system Hamiltonian as $\pmb{H}_S = \pmb{H}_S^{(0)} + \pmb{S}\langle\pmb{B}\rangle_B $ and the system--bath interaction as $\pmb{V} = \pmb{S}\otimes\pmb{B}-\pmb{S}\langle\pmb{B}\rangle_B$. This may be understood as a counter term to first order in $\zeta$ \cite{winczewski2021renormalization}.}. Note that this equation neglects any terms of order higher than $ \zeta^2 $.

Owing to the Born approximation, the global state can be assumed to factorise also at $t>0$ (i.e., $\state(t)\simeq\state_S(t)\otimes\pmb{\pi}_B$), while the Born--Markov approximation allows for the time-local structure of Eq.~\eqref{eq:Redfield-general}. Here, $\widetilde{\pmb O}(t)= e^{i\,\pmb{H}_{SB}\,t}\,\pmb{O}\,e^{-i\,\pmb{H}_{SB}\,t}$ denotes interaction picture with respect to $\pmb{H}_{SB} = \pmb{H}_S + \pmb{H}_B$, and $[\bullet,\bullet]$ is a commutator. In the absence of counter term, we simply set $\pmb{H}_S = \pmb{H}_S^{(0)}$; this will be our default position throughout the paper. However, when explicitly stated, $ \pmb{H}_S $ may also stand for $ \pmb{H}_S = \pmb{H}_S^{(0)} \pm \frac12\delta\omega^2\,\pmb{x}^2$ { (see Table~\ref{table1} below)}. 

We can separate the coherent contributions to the open dynamics by bringing Eq.~\eqref{eq:Redfield-general} into the form $d\widetilde{\state}_S/dt = -i\,[\widetilde{\Delta\pmb{H}}(t),\,\widetilde{\state}_S] + \widetilde{\pazocal{R}}(t)\,\widetilde{\state}_S$ (see, e.g., Refs.~\cite{cattaneo2019local,winczewski2021renormalization,timofeev2022hmf}), where the Hamiltonian correction $\widetilde{\Delta\pmb{H}}$ is
\begin{align}\label{eq:coherent-hamiltonian}
	\widetilde{\Delta\pmb{H}} = -\frac{i}{2}\,\zeta^2\,\int_0^\infty ds\,\tr_B\, \left([\widetilde{\pmb{V}}(t),\,\widetilde{\pmb{V}}(t-s)]\,\state_B\right);
\end{align}
this and the superoperator $\pazocal{R}$ are written more explicitly in Eqs.~\eqref{eq:Redfield-explicit-combo} below. Finally, moving to the Schr\"{o}dinger picture gives 
\begin{equation}\label{eq:redfield-abbreviated}
	\frac{d\state_S}{dt} = -i[\pmb H_S + \Delta\pmb H,\,\pmb{\varrho}_S] + \pazocal{R}\,\state_S(t),
\end{equation}

\subsection{Connection between Lamb shift and potential renormalisation}\label{sec:lamb-shift}

Let us now illustrate how, in the adiabatic regime, the Lamb shift encodes the expected distortion on the potential of the system. For now, we limit ourselves to our specific harmonic-oscillator model. However, in Sec.~\ref{sec:proof} we shall be in a position to show this in an arbitrary open system coupled, through a generic spectral density, to a harmonic bath at any temperature.

From Eq.~\eqref{eq:coherent-hamiltonian}, we find the effective correction to our system's Hamiltonian to be
\begin{equation}\label{eq:effective-Hamiltonian-oscillator}
	\Delta\pmb{H} = -\frac12\,\Sigma'\,\pmb{x}^2 + \frac{\Delta'}{4\,\omega_0} - \frac{\Delta}{8\,\omega_0}\,\{\pmb x,\,\pmb p\},
\end{equation}
where the notations $\Delta$, $\Delta'$ and $\Sigma'$ are \begin{subequations}\label{eq:delta-deltap-sigmap}
	\begin{align}
		\Delta &= -2\,J(\omega_0), \label{eq:delta}\\
		\Sigma' &= \frac{1}{\pi}~\dashint_{-\infty}^\infty d\omega \, \frac{J(\omega)}{\omega-\omega_0} { = \pazocal{H}[\,J(\omega)\,](\omega_0)}, \label{sigma-prime}\\
		\Delta' &= -{ \pazocal{H}\left[\,J(\omega)\,\left(2\,n(\omega)+1\right)\,\right](\omega_0)}, \label{eq:delta-prime}
	\end{align}
\end{subequations}
$n_\beta(\omega) = (e^{\beta\omega}-1)^{-1}$ is the bosonic occupation, { $\pazocal{H}[\,f(x)\,](y)$ for the Hilbert transform of $ f(x) $ evaluated at $ y $ \cite{bateman1954tables}}, and $\{\bullet,\bullet\}$ for anti-commutator. Importantly, the variables introduced in Eqs.~\eqref{eq:delta-deltap-sigmap} relate to the decay rate $\gamma(\omega)$ and Lamb shift\footnote{The term `Lamb shift' is usually reserved for $-\pazocal{H}\,J(\omega)$, while $-\pazocal{H}\,[J(\omega)\,n_\beta(\omega)]$ is referred-to as `Stark shift'. Here, however, we shall simply use `Lamb shift' to denote the sum of both.} $S(\omega)$ \cite{bp} 
\begin{align*}
	\gamma(\omega) &= 2\,J(\omega)\,(1+n_\beta(\omega)),\\
	S(\omega) &= -{ \pazocal{H}\big[\,J(\omega')\,\left( n_\beta(\omega')+1 \,\right)\big](\omega)}, 
\end{align*}
which are the real and imaginary parts of the Laplace transform of the bath correlation function $ \langle \widetilde{\pmb{B}}(t)\widetilde{\pmb{B}}(t-s) \rangle_B = \tr_B[\widetilde{\B}(t)\,\widetilde{\B}(t-s)\,\state_B] $
\begin{equation}\label{eq:complex-rates}
	\Gamma_\omega=\int_0^\infty ds\,e^{i\omega s}\,\langle \widetilde{\pmb{B}}(t)\widetilde{\pmb{B}}(t-s) \rangle_B = \frac12\,\gamma(\omega) + i\,S(\omega).
\end{equation}
Back to Eq.~\eqref{eq:delta-deltap-sigmap}, we can thus write 
\begin{subequations}
	\begin{align*}
		\Delta &= \gamma(-\omega_0) - \gamma(\omega_0) \\
		\Sigma' &= - S(-\omega_0) - S(\omega_0) \\
		\Delta' &= S(-\omega_0) - S(\omega_0).
	\end{align*}
\end{subequations}
Looking at the first two terms of Eq.~\eqref{eq:effective-Hamiltonian-oscillator}, we see that the Lamb shift effectively renormalises the bare harmonic potential (i.e., $\omega_0^2 \mapsto \omega_0^2 - \Sigma'$) \cite{ishizaki2009adequacy,hartmann2020environmentally}, and adds a uniform shift of the energy spectrum of the system. In addition, the correction $\Delta\pmb{H}$ contains a `squeezing term' proportional to $\{\pmb{x},\pmb{p}\}$ \cite{duffus2017open}; this is not related to $S(\omega)$. Interestingly, for our spectral density
\begin{equation*}
	\Sigma' = \frac{\lambda\Lambda}{1+(\omega_0/\Lambda)^2},
\end{equation*}
so that, in the adiabatic regime of $\omega_0/\Lambda\ll 1$, $\Sigma'\simeq\lambda\Lambda = \delta\omega^2$. That is, $-\frac12\,\Sigma'\pmb{x}^2$ does approach the reorganisation energy. 

The links between potential renormalisation and Lamb shift become even clearer when considering the stationary state of Eq.~\eqref{eq:redfield-abbreviated} for our model: at long times the system approaches the high-temperature limit $ \pmb{\tau}_{MF} $ in Eq.~\eqref{eq:mean-force-classical} if the $ S(\omega) $ terms are kept, and the local thermal state $ \pmb{\pi}_S $ if they are eliminated. Indeed, note that the steady state of the system must be a Gaussian, since the overall Hamiltonian is quadratic \cite{ferraro2005gaussian}. Hence, it can be fully characterised by the first- and second-order moments: $\tr_S[\pmb{x}\,\state_S(t\rightarrow\infty)]=\langle \pmb{x} \rangle_\infty = 0 $, $ \langle\pmb{p}\rangle_\infty = 0 $, $ \frac12\langle \{ \pmb{x},\,\pmb{p} \} \rangle_\infty = 0$, and 
\begin{subequations}\label{eq:steady-Redfield}
	\begin{align}
		\langle \pmb{x}^2 \rangle_\infty &= \frac{\omega_0}{2(\omega_0^2-\Sigma')}\left(\coth{\frac{\omega_0}{2T}} -\frac{\Delta'}  {\omega_0^2}\right),\label{eq:xx-redfield}\\ 
		\langle\pmb{p}^2\rangle_\infty &= \frac{\omega_0}{2}\,\coth{\frac{\omega_0}{2T}}\label{eq:pp-redfield},
	\end{align}
\end{subequations}
with $\Sigma = -\gamma(-\omega_0)-\gamma(\omega_0)$ (cf. Appendix~\ref{app:redfield}). We can now easily check that, at large temperatures, the fidelity between the state defined by Eqs.~\eqref{eq:steady-Redfield} and the correct classical limit of the mean-force Gibbs state $\pmb{\tau}_{MF}$ from Eq.~\eqref{eq:general-high-T-mean-force} is indeed high. However, if $ S(\omega) $ is set to zero (and thus $\Sigma'=\Delta'=0$), and we are left with $\langle\pmb{x}^2\rangle_\infty = \coth{(\omega_0/2T)}/2\omega_0$, which corresponds the local equilibrium state $\pmb{\pi}_S$, regardless of $T$ or $\lambda$.

\subsection{Secular approximation}\label{sec:secular-approximation}

We have thus established that the Lamb-shift terms carry information about the distortion of the system's potential. Now we can be more specific and pinpoint \textit{which} Lamb-shift terms within Eq.~\eqref{eq:redfield-abbreviated} matter in terms of potential renormalisation. To do so, let us cast the master equation in a more explicit form. Namely, we decompose the system--bath coupling operator as $ \pmb{S} = \sum\nolimits_\omega \pmb{A}_\omega $. Here, the sum runs over all Bohr frequencies of $ \pmb{H}_S $ and the non-Hermitian operators $\pmb{A}_\omega$ are constructed so that $[\pmb{H}_S,\,\pmb{A}_\omega] = - \omega\,\pmb{A}_\omega$ and $ \pmb{A}_{-\omega} = \pmb{A}_\omega^\dagger $. For instance, in our oscillator model, these would be $\pmb{A}_{\omega} = \frac12(\pmb{x}+\frac{i}{\omega}\,\pmb{p})$ and $\omega\in\{\omega_0,-\omega_0\}$. The Redfield equation \eqref{eq:redfield-abbreviated} then reads
\begin{subequations}\label{eq:Redfield-explicit-combo}
	\begin{equation}\label{eq:Redfield-explicit}
		\frac{d\pmb{\varrho}_S}{dt} = -i\,[\pmb{H}_S+\Delta\pmb{H},\pmb{\varrho}_S] + \sum_{\omega,\omega'}\Gamma_{\omega}\left( \pmb{A}_\omega\,\pmb{\varrho}_S\,\pmb{A}_{\omega'}^\dagger - \frac12\,\{\pmb{A}_{\omega'}^\dagger\,\pmb{A}_\omega,\,\pmb{\varrho}_S\} \right) + \text{h.c.},
	\end{equation}
	where the complex `rates' $\Gamma_\omega = \frac12\,\gamma(\omega) + i\,S(\omega)$ were defined in Eq.~\eqref{eq:complex-rates}. {  Note that both real and imaginary part scale as $ \zeta^2 $}. The term $\Delta\pmb{H}$ [cf. Eq.~\eqref{eq:coherent-hamiltonian}] takes the form
	\begin{equation}\label{eq:coherent-hamiltonian-explicit}
		\Delta \pmb{H} = -\frac{i}{2}\sum\nolimits_{\omega,\omega'} \Gamma_{\omega}\, \pmb{A}_{\omega'}^\dagger\,\pmb{A}_\omega + \text{h.c.}
	\end{equation}
\end{subequations}

We can further perform the \textit{secular approximation} on Eqs.~\eqref{eq:Redfield-explicit-combo}, which consists in eliminating all terms with $\omega\neq\omega'$. Owing to the definition of the operators $\pmb{A}_\omega$, these terms oscillate with frequency $\vert\omega-\omega'\vert$ in the interaction picture. Provided that these oscillations are sufficiently fast compared with the dissipation timescale $ t_D \sim \lambda^{-1} $, their elimination may be justified as the result of coarse-graining \cite{bp}. Importantly, after the secular approximation Eq.~\eqref{eq:Redfield-explicit} adopts the familiar GKLS form,
\begin{equation}\label{eq:Lindblad}
	\frac{d\pmb{\varrho}_S}{dt} = -i\left[\pmb{H}_S + \sum\nolimits_{\omega} S(\omega)\,\pmb{A}_\omega^\dagger\,\pmb{A}_\omega,\,\pmb{\varrho}_S\right] + \sum\nolimits_{\omega} \gamma(\omega)\,\left( \pmb{A}_\omega\,\pmb{\varrho}_S\,\pmb{A}_\omega^\dagger - \frac12\,\{ \pmb{A}_\omega^\dagger\,\pmb{A}_\omega,\,\pmb{\varrho}_S \} \right),
\end{equation}
and thus enjoys complete positivity. 

Note how some of the Lamb-shift terms do survive the secular approximation and yet, the resulting steady state is $\pmb{\varrho}_S(t\rightarrow\infty) = e^{-\beta\,\pmb{H}_S}/\tr_S{e^{-\beta\,\pmb{H}_S}} = \pmb{\pi}_S$ as expected for a GKLS equation \cite{bp}. Hence, it is the `non-secular' Lamb-shift terms which capture the distortion of the system's potential. 

{ The GKLS equation \eqref{eq:Lindblad} is generally regarded as preferable over the Redfield equation \eqref{eq:Redfield-explicit-combo}---especially in quantum thermodynamics \cite{alicki1979engine,dann2021compatibility}---since the non-secular terms of the latter may break complete positivity during the transient dynamics \cite{gaspard1999slippage,hartmann2020embracing}. In contrast, as already advanced, completely positive evolution is guaranteed under a GKLS generator. Our example, however, illustrates that the positivity of the textbook GKLS equation comes at price---approaching the wrong classical limit. As we will show in Sec.~\ref{sec:proof}, this can be remedied by a consistent perturbative treament of the reorganisation energy in the derivation of the master equation, which ammounts to adopting the recipe outlined in Sec.~\ref{sec:intro}.}

\subsection{To shift or not to shift?}\label{sec:to-shift-or-not-to-shift}

\subsubsection{When there is no counter term}

To bring the issue of the removal of the Lamb shift into focus, let us go back to the Redfield equation \eqref{eq:Redfield-explicit-combo} and its steady state \eqref{eq:steady-Redfield}. As already mentioned, Eqs.~\eqref{eq:steady-Redfield} approach the classical limit \eqref{eq:mean-force-classical} at high temperatures, albeit not exactly \cite{thingna2012generalized}. Indeed, even if the bath were tuned so that $ \Sigma' = \delta\omega^2 $, the expected negative frequency shift would not appear in the expression for $\langle\pmb{p}^2\rangle_\infty$, regardless of the temperature (cf. Eq.~\eqref{eq:pp-redfield}). On the other hand, we know that, in the absence of Lamb shift, Eqs.~\eqref{eq:steady-Redfield} match the local Gibbs state with respect to the \textit{bare} Hamiltonian $\pmb{H}_S^{(0)}$. But, what if we were to shift the frequency of the oscillator \textit{down}, so that
\begin{equation*}
	\pmb{H}_S \mapsto \pmb{H}_R = \pmb{H}_S^{(0)}-\frac12\delta\omega^2\,\pmb{x}^2,
\end{equation*}
whilst removing all Lamb-shift terms `by hand' in the resulting Redfield equation? { That is, setting $ S(\omega) = 0 $, which cancels all Lamb-shift contributions to the Hamiltonian correction as well as to the dissipative part of the master equation.} The asymptotic state would then be
\begin{equation}\label{eq:semi-classical-state}
	\pmb{\varrho}_S(t\rightarrow\infty) = \frac{e^{-\beta\left[\frac12(\omega_0^2-\delta\omega^2)\,\pmb{x}^2 + \frac12\,\pmb{p}^2\right]}}{\tr_S\,e^{-\beta\left[\frac12(\omega_0^2-\delta\omega^2)\,\pmb{x}^2 + \frac12\,\pmb{p}^2\right]}},
\end{equation}
which is precisely $\pmb{\tau}_{MF}(\beta\ll\omega_0)$. \textit{A priori}, we would expect such equation to be accurate only at very high temperatures---i.e., in the classical regime---and long times. However, as we illustrate below, it remains asymptotically accurate even when far from the classical limit (cf. Fig.~\ref{fig1}). That is, eliminating the Lamb-shift terms proves to be an effective artefact \textit{if accompanied by the right manipulations on the system Hamiltonian}---namely, subtracting the reorganisation energy term. { Furthermore, as we shall see in Sec.~\ref{sec:proof-3} below, this modified equation accurately captures the frequency of transient oscillations of arbitrary open systems in the adiabatic regime.} 

In fact, the idea of building the correct long-time limit into the weak-coupling master equation, which will be central for us in what follows, had been considered previously \cite{thingna2012generalized,thingna2013reduced,lobejko2022towards,timofeev2022hmf,becker2022canonically}. Namely, the fitness of a Redfield equation constructed from approximations to the `mean-force Hamiltonian' (which satisfies $\pmb{\tau}_{MF} = \frac{e^{-\beta\,\pmb{H}_{MF}}}{\pazocal{Z}_{MF}}$ with $\pazocal{Z}_{MF} = \tr_S{e^{-\beta\,\pmb{H}_{MF}}}$) had been numerically studied, showing some improvement at long times \cite{timofeev2022hmf}. The reason for this success can be elegantly explained by considering the Hamiltonian renormalisation procedure that would bring the interaction-picture cumulant equation into a dissipative-only form \cite{lobejko2022towards,winczewski2021renormalization}. Similarly, using a Redfield equation modified to steer the system towards the mean-force Gibbs state has been shown to outperform conventional weak-coupling equations at long times, and even to improve the accuracy of the transient dynamics of a damped harmonic oscillator \cite{becker2022canonically}.

\subsubsection{When the counter term is present}

Importantly, Born--Markov master equations are often used in the context of quantum optics \cite{carmichael2009open}, thus modelling coupling to the electromagnetic field. In this case, the {  microscopic Hamiltonian} must explicitly include a counter term compensating for any potential renormalisation \cite{caldeira1983quantum}. As already advanced, $ \pmb{H}_S = \pmb{H}_S^{(0)} + \frac12\delta\omega^2\,\pmb{x}^2 $ would then play the role of the system Hamiltonian { (see Table~\ref{table1})}. Strictly speaking, it is this $\pmb{H}_S$---counter term included---which must be used when deriving the Redfield equation. And, in principle, we should keep all the resulting Lamb-shift terms. In practice, however, the Hamiltonian $\pmb{H}_R = \pmb{H}_S^{(0)} = \pmb{H}_S - \frac12\delta\omega^2\,\pmb{x}^2$ is almost universally used, while all Lamb-shift terms are removed from the master equation. Doing so, one arrives to local equilibrium with respect to the bare Hamiltonian $\pmb{H}_S^{(0)}$. That is, one recovers the correct classical limit and, more generally, a good approximation to $\pmb{\tau}_{MF}$. { That is, the common practice in this case is exactly the same: one subtracts the reorganisation energy from the the \textit{microscopic} Hamiltonian $\pmb{H}_S$---which cancels the counter term---while the Lamb shift is eliminated.} 

\begin{table}
{ 
\begin{center}
\begin{tabular}{ | >{\centering\arraybackslash}p{4cm} | >{\centering\arraybackslash}p{4cm} | >{\centering\arraybackslash}p{4cm} |  }
 \cline{2-3}
\multicolumn{1}{c|}{} & Microscopic Hamiltonian ($\pmb{H}_S$) & Renormalised Hamiltonian ($\pmb{H}_R$)\\
 \hline
 Includes counter term   & \mline{$\pmb{H}_S^{(0)} + Q\,\pmb{S}^2$}    & \mline{$\pmb{H}_S^{(0)}$}\\
 \hline
 No counter term   & \mline{$\pmb{H}_S^{(0)}$}    & \mline{$\pmb{H}_S^{(0)} - Q\,\pmb{S}^2$}\\
 \hline
\end{tabular}
\end{center}
\caption{ {  Microscopic Hamiltonian}s for an open system with and without counter-term, and their renormalised versions entering the derivation of a reorganised master equation.}
\label{table1}
}
\end{table}

\section{Revisiting the Redfield equation}\label{sec:proof}

\subsection{Steps (i) and (ii)}\label{sec:proof-12}

We now discuss the precise conditions in which the common practice may be rigorously justified for \textit{arbitrary} open systems. We begin by looking at the first two steps summarised in Sec.~\ref{sec:intro}; that is, to subtract the reorganisation energy from $ \pmb{H}_S $ and obtain a master equation from the resulting Hamiltonian. All we need to assume is a generic form for the spectral density; namely,
\begin{align}\label{eq:genera-spectral-density}
     J(\omega) = \lambda\,\Lambda\,(\omega/\Lambda)^s\,f(\omega/\Lambda),
\end{align}
where $ s $ is the Ohmicity parameter and $f(x)$ is a smooth function satisfying $ f(0) \neq 0 $ and decaying sufficiently fast for $ x > 1 $. Note that this includes the Ohmic-algebraic case from Eq.~\eqref{eq:spectral_density_OhmicAlg} or the common choice $ J(\omega) = \lambda\,\Lambda\,(\omega/\Lambda)^s\,e^{-\omega/\Lambda}$ with variable Ohmicity $s$.

Let us go back to the interaction-picture form of the second-order Born--Markov master equation \eqref{eq:Redfield-general}. { We are going to choose $ \pmb{H}_R = \pmb{H}_S - Q\,\pmb{S}^2$ as our zeroth-order Hamiltonian, so that the {  microscopic Hamiltonian} $\pmb{H}_S = \pmb{H}_R + Q\,\pmb{S}^2 $ is just a small perturbation away. This justifies a series expansion of the interaction-picture form of the system--bath coupling operator $ \pmb{H}_\text{int}=\zeta\,\pmb{V} $ as follows:
\begin{multline}\label{eq:2nd-order-interaction-picture}
    \zeta\,\widetilde{\pmb{V}}(t) = \zeta\,e^{i\,(\pmb{H}_R + Q\,\pmb{S}^2 + \pmb{H}_B)\,t}\,\Hint\,e^{-i\,(\pmb{H}_R + Q\,\pmb{S}^2 + \pmb{H}_B)\,t} \\= \zeta\,e^{i\,(\pmb{H}_R + \pmb{H}_B)\,t}\,\Hint\,e^{-i\,(\pmb{H}_R + \pmb{H}_B)\,t}  + \pazocal{O}(Q\,\zeta). 
\end{multline}
Noting that $ Q = \frac12\,\pazocal{H}\,J(0) $ scales with $ \lambda $ (cf. Eq.~\eqref{eq:genera-spectral-density}), which is $\pazocal{O}(\zeta^2)$, we can see that the largest terms that we have dropped is actually $\pazocal{O}(\zeta^3)$. Terms that small may indeed be safely removed from a second-order master equation}. The resulting expression
\begin{equation}\label{eq:true-2ndorder-redfield}
	\frac{d\widetilde{\pmb\varrho}_S}{dt} = - \zeta^2\int_0^\infty ds\,\tr_B\,[\widetilde{\pmb{V}}_0(t),\,[\widetilde{\pmb{V}}_0(t-s),\,\widetilde{\pmb\varrho}_S(t)\otimes\pmb{\pi}_B]],
\end{equation}
where we have defined
{ 
\begin{equation*}
    \widetilde{\pmb{V}}_ 0(t) = e^{i\,(\pmb{H}_R + \pmb{H}_B)\,t}\,\Hint\,e^{-i\,(\pmb{H}_R + \pmb{H}_B)\,t},
\end{equation*}
}
{ is as accurate as Eq.~\eqref{eq:Redfield-general}}, in the sense that both neglect terms smaller than $\pazocal{O}(\zeta^2)$. 

To further simplify Eq.~\eqref{eq:true-2ndorder-redfield} it is convenient to decompose $\pmb{S}$ in eigenoperators $\pmb{A}^{(0)}_\nu$ of the { zeroth-order Hamiltonian $ \pmb{H}_R = \pmb{H}_S - Q\,\pmb{S}^2 $, such that $[\pmb{H}_R,\pmb{A}_{\nu}^{(0)}] = -\nu\,\pmb{A}_{\nu}^{(0)}$}. Here, we use the notation $ \nu $ for the Bohr frequencies of $ \pmb{H}_R $. The resulting equation would thus be
\begin{align}\label{eq:true-redfield-interaction}
    \frac{d\widetilde{\pmb{\varrho}}_S}{dt} = -i\,[\widetilde{\Delta\pmb{H}}_0(t),\pmb{\varrho}_S] + \sum\nolimits_{\nu,\nu'}\Gamma_{\nu}\left( \pmb{A}^{(0)}_\nu\,\pmb{\varrho}_S\,{\pmb{A}^{(0)}_{\nu'}}^\dagger -\frac12\,\{{\pmb{A}^{(0)}_{\nu'}}^\dagger\,\pmb{A}^{(0)}_\nu,\,\pmb{\varrho}_S\} \right)\,e^{-i(\nu-\nu')t} + \text{h.c.},
\end{align}
where the Hamiltonian-like term is now
\begin{equation*}
    \widetilde{\Delta \pmb{H}}_0(t) = -\frac{i}{2}\sum\nolimits_{\nu,\nu'} \Gamma_{\nu}\, {\pmb{A}_{\nu'}^{(0)}}^\dagger\,\pmb{A}^{(0)}_\nu\,e^{-i(\nu-\nu')t} + \text{h.c.}
\end{equation*}
{ When bringing Eq.~\eqref{eq:true-redfield-interaction} back into the Schr\"{o}dinger picture, we invoke the same argument as in Eq.~\eqref{eq:2nd-order-interaction-picture}; that is, 
\begin{equation*}
    e^{-i(\pmb{H}_S + \pmb{H}_B)t}\,\widetilde{\pmb{A}_\nu^{(0)}}(t)\,e^{i(\pmb{H}_S + \pmb{H}_B)t} = e^{i\,(\pmb{H}_R + \pmb{H}_B)\,t}\,\widetilde{\pmb{A}_\nu^{(0)}}(t)\,e^{-i\,(\pmb{H}_R + \pmb{H}_B)\,t} + \pazocal{O}(\zeta^2) = \pmb{A}_\nu^{(0)}+ \pazocal{O}(\zeta^2),
\end{equation*}
where $ \widetilde{\pmb{A}_\nu^{(0)}}(t) = \pmb{A}_\nu^{(0)}\,e^{-i\,\nu\,t}$. Crucially, recalling that both real and imaginary parts of $ \Gamma_\nu $ are already $ \pazocal{O}(\zeta^2) $ we readily get
\begin{equation}\label{eq:true-Redfield-non-adiabatic}
    \frac{d\pmb{\varrho}_S}{dt} = -i\,\left[\pmb{H}_S-\frac{i}{2}\sum_{\nu,\nu'} \Gamma_{\nu}\, {\pmb{A}_{\nu'}^{(0)}}^\dagger\,\pmb{A}^{(0)}_\nu,\pmb{\varrho}_S \right] + \sum_{\nu,\nu'}\Gamma_{\nu}\left( \pmb{A}^{(0)}_\nu\,\pmb{\varrho}_S\,{\pmb{A}_{\nu'}^{(0)}}^\dagger - \frac12\,\{{\pmb{A}_{\nu'}^{(0)}}^\dagger\,\pmb{A}^{(0)}_\nu,\,\pmb{\varrho}_S\} \right) + \text{h.c.},
\end{equation}
which, once again, can be claimed to be accurate up to $\pazocal{O}(\zeta^2)$.}

The key difference between Eq.~\eqref{eq:true-Redfield-non-adiabatic} and other second-order master equations is that all dissipative processes in \eqref{eq:true-Redfield-non-adiabatic} occur in the basis of the relevant Hamiltonian $\pmb{H}_S-Q\,\pmb{S}^2$ which defines the classical limit of the system's true stationary state (cf. Eq.~\eqref{eq:mean-force-classical}). In practice, for Eq.~\eqref{eq:true-Redfield-non-adiabatic} to be valid, one needs to ensure that $ Q $ is sufficiently small for the chosen parameters. Note that, for our specific example $ Q = \lambda\,\Lambda$, so that $ Q $ might be large even for weak $ \lambda $. However, if $ \Lambda \ll \frac{\omega_0}{\lambda} $ Eq.~\eqref{eq:true-Redfield-non-adiabatic} may be applied safely.

We have thus shown that, as long as $ Q $ is small, steps (i) and (ii) lead to the same second-order master equation as a rigorous microscopic derivation. 

\subsection{Step (iii)}\label{sec:proof-3}

We are finally in a position to tackle the cancellation of the Lamb shift and the reorganisation energy in the adiabatic regime. The explicit calculation of the Lamb shift yields 
\begin{equation*}
    S(\nu) = -\frac{1}{\pi}\,\dashint_{-\infty}^\infty d\nu'\,\frac{J(\nu')\,\big(n_\beta(\nu') + 1\big)}{\nu'-\nu} = -\frac{1}{\pi}\,\dashint_{-\infty}^\infty d\varphi\,\frac{J(\varphi)\,\big(n_{\beta\Lambda}(\varphi) + 1\big)}{\varphi-\omega/\Lambda},
\end{equation*}
where $\varphi = \nu'/\Lambda$. In the adiabatic regime, one has $ \nu/\Lambda \ll 1 $ for all the Bohr frequencies $\nu$ appearing in the double summation of \eqref{eq:true-Redfield-non-adiabatic}, meaning that
\begin{equation*}
    S(\nu) \simeq S(0) = -\frac{1}{\pi}\int_{0}^\infty d\varphi\,\frac{J(\varphi)}{\varphi} = -Q. 
\end{equation*}
Here, we have used the identity $n_\beta(-\nu)=-(1+n_\beta(\nu))$ and the fact that $J(-\nu)=-J(\nu)$. Note that we make no assumptions about the temperature $\beta$. Finally, performing the substitution $ S(\nu)\mapsto S(0) $ in Eq.~\eqref{eq:true-Redfield-non-adiabatic} gives
\begin{multline}\label{eq:artefact}
    \frac{d\pmb{\varrho}_S}{dt} \simeq -i[\pmb{H}_S - Q\,\pmb{S}^2,\pmb{\varrho}_S] -\frac14 \big[\sum\nolimits_{\nu,\nu'}\left(\gamma(\nu)-\gamma(\nu')\right)\,{\pmb{A}_{\nu'}^{(0)}}^\dagger\,\pmb{A}^{(0)}_{\nu} ,\pmb\rho_S\big]  \\ + \frac12\,\sum\nolimits_{\nu,\nu'} \left(\gamma(\nu)+\gamma(\nu')\right)\,\left( \pmb{A}^{(0)}_\nu\,\pmb{\varrho}_S\,{\pmb{A}_{\nu'}^{(0)}}^\dagger - \frac12\,\{{\pmb{A}_{\nu'}^{(0)}}^\dagger\,\pmb{A}^{(0)}_\nu,\,\pmb{\varrho}_S\} \right) + \,\text{h.c.}
\end{multline} 

We thus see that all Lamb-shift contributions collapse into $ i[Q\,\pmb{S}^2,\pmb{\varrho}_S] $, which effectively renormalises the {  microscopic Hamiltonian} in the expected way \cite{hartmann2020environmentally}. Note that the additional commutator-like term does not depend on the Lamb shift, but only the decay rates $ \gamma(\nu) $, and vanishes under the secular approximation. In our example model, this amounts for the aforementioned squeezing term, proportional to $\{\pmb{x},\pmb{p}\}$ (cf. Eq.~\eqref{eq:effective-Hamiltonian-oscillator}). Importantly, the fact that the coherent part of the master equation does indeed become $-i[\pmb{H}_S-Q\,\pmb{S}^2,\pmb{\varrho}_S]$ in the adiabatic regime (and for small $Q$), means that steps (i)--(iii) correctly capture, not only the classical limit of the steady state of arbitrary open systems, but also their transient oscillations. Also note that the most-often used secular-approximated version of Eq.~\eqref{eq:artefact} is guaranteed to converge asymptotically to the correct classical limit regardless of the magnitude of $ Q $ (cf. Eq.~\eqref{eq:mean-force-classical}). 

We have thus rigorously shown that renormalising the Hamiltonian of an open system by subtracting the reorganisation energy, and removing all Lamb shift terms when deriving a second-order master equation, may be admissible whenever $Q$ is small and $ \nu \ll \Lambda $ for all the relevant Bohr frequencies $\nu$ of the renormalised Hamiltonian { $\pmb{H}_R = \pmb{H}_S-Q\,\pmb{S}^2$}. This is our main result. In our example model these conditions write compactly as
\begin{equation*}
    \omega_0\ll\Lambda\ll\frac{\omega_0}{\lambda}.
\end{equation*} 

\section{Exact dynamics}\label{sec:exact}

Before illustrating our findings, let us briefly outline the exact solution of our example model \cite{lampo2019quantum}, that we use as a benchmark against which to compare the various forms of second-order master equations (full details are given in Appendix~\ref{app:Heisenberg}). The exact Heisenberg equations of motion for the system can be compacted into the well-known quantum Langevin equation \cite{weiss1999} 
\begin{align}
	\ddot{\pmb x}(t) + \omega_0^2\,\pmb x(t) - \int_0^t d\tau\,\chi(t-\tau)\, \pmb x(\tau) = \pmb F(t), \label{eq:QLE}
\end{align}
where $\chi(t)$ is the `dissipation kernel'
\begin{equation*}
	\chi(t) = \sum\nolimits_\mu \frac{g_\mu^2}{m_\mu\,\omega_\mu}\,\sin{(\omega_\mu\,t)} = \frac{2}{\pi} \int_0^\infty J(\omega) \sin{(\omega t)}\,d \omega,
\end{equation*}
and $\pmb{F}(t)$ is a quantum stochastic force, which encodes the initial conditions of the bath. 

Introducing the compact notation $ \pmb{\mathsf{Z}} = (\pmb x, \pmb p)^\mathsf{T} $ allows to write the exact solution of Eq.~\eqref{eq:QLE} as
\begin{subequations}\label{eq:G-t}
	\begin{equation}
		\pmb{\mathsf{Z}}(t) = \mathsf{G}(t)\,\pmb{\mathsf{Z}}(0) + \int_0^t \mathsf{G}(t-t')\,\pmb{\mathsf{F}}(t')\,dt',
	\end{equation}
	where the entries of $\mathsf{G}(t)$ are 
	\begin{align}
		[\mathsf{G}(t)]_{11} &= [\mathsf{G}(t)]_{22} =\pazocal{L}_t^{-1}\left[\frac{s}{s^2 + \omega_0^2 - \widehat{\chi}(s)}\right],\\
		[\mathsf{G}(t)]_{12} &=\pazocal{L}_t^{-1}\left[\frac{1}{s^2 + \omega_0^2 - \widehat{\chi}(s)}\right]= g(t),\\
		[\mathsf{G}(t)]_{21} &=\pazocal{L}_t^{-1}\left[\frac{\widehat{\chi}(s)-\omega_0^2}{s^2 + \omega_0^2 - \widehat{\chi}(s)}\right],
	\end{align}
\end{subequations}
and $\pmb{\mathsf{F}}$ is defined in Appendix~\ref{app:Heisenberg}. Here, $\widehat{f}(s):=\int_0^\infty dt\,e^{-s\,t}\,f(t)$ stands for Laplace transform, while we denote the inverse transform by $\pazocal{L}^{-1}_t[\,g\,]$. For our spectral density, $\widehat{\chi}(s)$ is
\begin{equation}\label{eq:covariance-matrix-general}
	\widehat{\chi}(s) =  \frac{\lambda \Lambda^2}{s + \Lambda} .
\end{equation} 
With this, we calculate the time evolution of the second-order moments (cf. Appendix~\ref{app:Heisenberg}), which can be grouped into the covariance matrix
\begin{equation}
	\mathsf{\Sigma}(t) = \textrm{Re}\,\langle \pmb{\mathsf{Z}}(t)\,\pmb{\mathsf{Z}}(t)^\mathsf{T} \rangle - \langle \pmb{\mathsf{Z}}(t)\rangle \langle \pmb{\mathsf{Z}}(t)^\mathsf{T}\rangle. \label{covmatrix}
\end{equation}

Specifically, in the long-time limit we have
\begin{subequations}\label{eq:covariances_coefficients}
	\begin{align}
		&\langle \pmb{x}^2 \rangle_\infty =\frac{1}{2}\int_0^\infty \vert \widehat{g}(i \omega )\vert^2\,\nu(\omega)\,d\omega, \label{eq:xx_ss}\\
		&\langle \pmb{p}^2 \rangle_\infty =\frac{1}{2}\int_0^\infty \omega^2\,\vert\widehat{g}(i \omega )\vert^2\,\nu(\omega)\,d\omega,  \label{eq:pp_ss}\\
		&\langle \{ \pmb{x},\,\pmb{p} \} \rangle_\infty =0 \label{eq:xp_ss}, 
	\end{align}
\end{subequations}
where we have introduced the notation
\begin{equation}
	\nu(\omega) = \frac{2}{\pi}\,J(\omega)\,\coth{\frac{\omega}{2T}}. \label{eq:nk}
\end{equation}
Eqs.~\eqref{eq:covariances_coefficients} thus yield the exact { state}. More generally, Eqs.~\eqref{eq:G-t} and \eqref{eq:covariances_coefficients} above give the exact dynamics and steady state of any arbitrary harmonic network of harmonic oscillators, once decoupled into normal modes \cite{martinez2013linear_thermo,freitas2014linear_analytic}.

\section{Illustration and discussion}\label{sec:results}

\begin{figure*}[t]
	\centering
	\includegraphics[width=\textwidth]{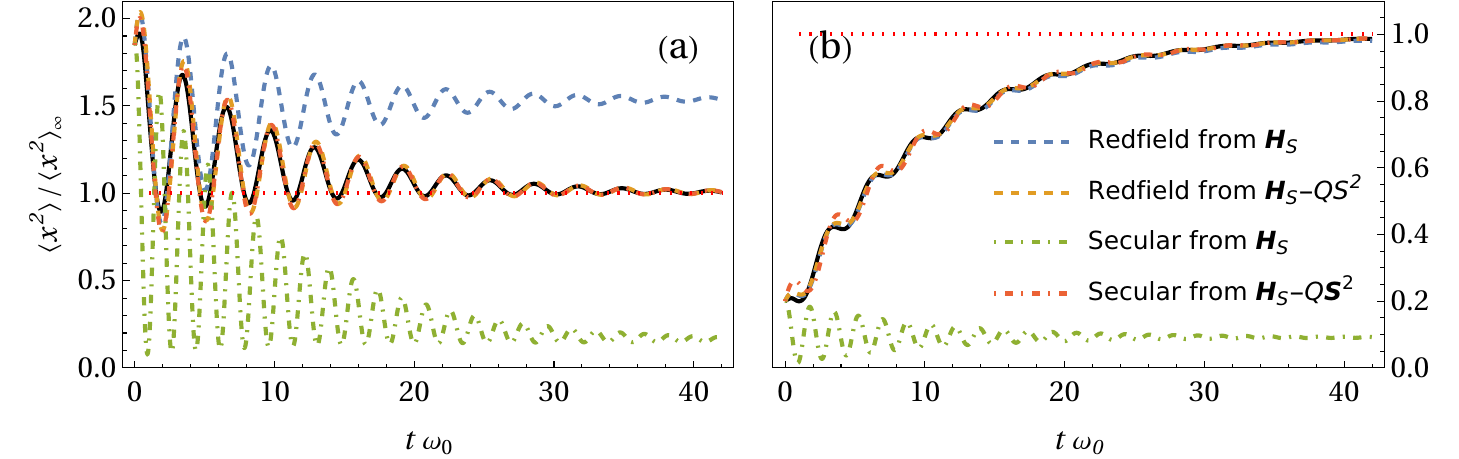}
	\caption{\textbf{Dynamics of a damped harmonic oscillator.} Exact thermalising dynamics of the position variance $\textstyle{\langle \pmb{x}^2(t) \rangle}$ relative to $\langle\pmb{x}^2\rangle_\infty$, for an oscillator with Hamiltonian $ \textstyle{\pmb{H}_S = \pmb{H}_S^{(0)} + \frac12\delta\omega^2\,\pmb{x}^2} $ (solid black) at an intermediate coupling of $\lambda = 0.1\,\omega_0$. The Redfield equation \eqref{eq:Redfield-explicit-combo} derived from this Hamiltonian (dashed Blue) is accurate in the classical regime, as shown in panel (b), but breaks down at moderate and low temperatures (see (a)). Further performing the secular approximation yields even worse results (dot-dashed green). Specifically, the resulting GKLS equation forces the system into a thermal state with respect to $ \textstyle{\pmb{H}_S = \pmb{H}_S^{(0)} + \frac12\delta\omega^2\,\pmb{x}^2} $, which is incompatible with the correct high-temperature limit from Eq.~\eqref{eq:semi-classical-state}. On the contrary, a Redfield equation derived from $ \textstyle{\pmb{H}_S = \pmb{H}_S^{(0)}} $, where the counter term is removed by an adding the reorganisation energy, yields excellent results provided that all Lamb-shift terms are {eliminated} (dashed yellow). The secular version of this modified equation also tracks the dynamics very reliably (dot-dashed orange). Both of these equations succeed in bringing the system close to the correct mean-force Gibbs state (red dotted line) over a very broad temperature range. The parameters are $\omega_0=1, \lambda=0.1$ and $\Lambda=100$. Hence, $ \delta\omega^2 = 10 $. { The initial state chosen is $\langle\pmb x(0)\rangle = \langle\pmb{p}^2(0)\rangle = 1$, $\langle\pmb p(0)\rangle = \langle\pmb{x}^2(0)\rangle = 2$, $\frac12\langle \{\pmb{x}(0),\pmb{p}(0)\} \rangle=\frac12$}. Recall that we work in units of 
		$ \hbar = k = 1 $ . In panel (a) $T=1$ and in (b) $T=10$.}
	\label{fig1}
	\centering
\end{figure*}

\begin{figure*}[t]
	\centering
	\includegraphics[width=\textwidth]{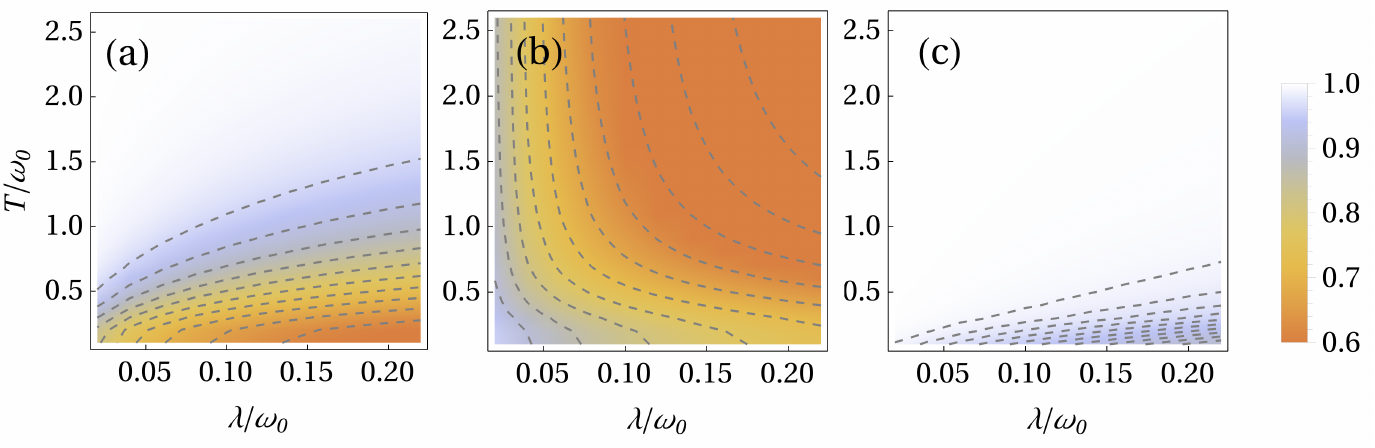}
	\caption{\textbf{Benchmarking the steady state of the master equations.} Uhlmann fidelity between the exact steady state $\pmb{\tau}_{MF}$ and the asymptotic state resulting from different master equations, for varying temperature and system--bath coupling. Just like in Fig.~\ref{fig1} the microscopic model includes an explicit counter term. In (a) the resulting full Redfield equation---Lamb shift included---is benchmarked against the mean-force Gibbs state. As we see, the agreement is good as long as the temperature is not too low, or the coupling too strong. In (b) we performed the secular approximation, which yields the thermal state corresponding to $ \pmb{H}_S = \pmb{H}_S^{(0))} + Q\,\pmb{S}^2 $ asymptotically. This is a very bad approximation to $\pmb{\tau}_{MF}$ due to the large reorganisation energy in our example. Finally, in (c) we cancel the counter term and vanish the Lamb shift. The resulting steady state is the high-temperature limit from Eq.~\eqref{eq:semi-classical-state}, which turns out to be an excellent approximation to the mean-force Gibbs, other than at very low temperatures.}
	\label{fig2}
	\centering
\end{figure*}

\subsection{Dynamics}

Let us start by looking at the transient dynamics of our model. In particular, we shall assume that the {  microscopic Hamiltonian} has a built-in counter term, i.e., 
\begin{align*}
	\pmb{H}_S = \pmb{H}_S^{(0)} + \frac12\delta\omega^2\,\pmb{x}^2,
\end{align*}
and will focus on the evolution of the position variance $\langle\pmb{x}^2(t)\rangle$. Its exact dynamics, as per Eq.~\eqref{eq:G-t}, is plotted on solid black in Fig.~\ref{fig1}. We also depict the evolution according to the Redfield equation \eqref{eq:Redfield-explicit-combo} derived from the above Hamiltonian (dashed blue). As already advanced, at high temperatures the agreement is good (see Fig.~\hyperref[fig1]{1(b)}). In contrast, at low and even \textit{moderate} temperatures---i.e., so that $T \lesssim \omega_0 $---there may be a significant error (cf. Fig.~\hyperref[fig1]{1(a)}). We note however, that this deviation might be partly attributed to the rather large coupling strength of $\lambda = 0.1\,\omega_0$. If the secular approximation is applied on such Redfield equation, the resulting dynamics (depicted in dot-dashed green) deviates from the exact one even further---neither the frequency of the oscillations nor the stationary value are captured correctly, regardless of temperature. Indeed, we know that the stationary state of Eq.~\eqref{eq:Lindblad} is always thermal with respect to $ \pmb{H}_S = \pmb{H}_S^{(0)} + \frac12\delta\omega^2\,\pmb{x}^2 $. The much narrower position distribution predicted by the secular equation in the steady state is thus explained by our choice of parameters, for which $\delta\omega^2/\omega_0^2 = 10 $.

In contrast, if we proceed as discussed in Sec.~\ref{sec:to-shift-or-not-to-shift}, i.e., by cancelling the counter term with the reorganisation energy---so that { $ \pmb{H}_S \mapsto \pmb{H}_R = \pmb{H}_S^{(0)} $}---and set $S(\omega)=0$ in Eq.~\eqref{eq:Redfield-explicit-combo}, the resulting Redfield equation \eqref{eq:artefact} showcases an excellent agreement with the exact solution, especially at $t\gg t_D$ (see dashed yellow curve). Further applying the secular approximation on this equation, as most commonly done in the literature, still matches the exact results to a very good approximation (see dot-dashed orange curve).

We remark that the choice of parameters for our illustrations lies well beyond the safe range discussed in Sec.~\ref{sec:proof-12}. In particular, we do work in the adiabatic regime of $\omega_0 \ll \Lambda$, but the reorganisation energy term is far from a perturbation for the chosen coupling strength. This was done simply to aid visualisation, as it increases the differences between steady states according to the various equations. The fact that Eq.~\eqref{eq:artefact} remains, nonetheless, so accurate is likely a model-specific feature. However, this numerical observation indicates that Eq.~\eqref{eq:artefact} still yields the correct stationary state in the classical limit, as well as at virtually any temperature (see Fig.~\ref{fig2}), and the correct transient oscillatory dynamics, for arbitrary harmonic networks of harmonic oscillators, even when the reorganisation energy is comparatively large. This is because an all-harmonic network may be always decomposed in normal modes and thus, reduced to our toy model. What can be said in general, however, is that the common practice is always accurate at long times and in the high-temperature limit, \textit{regardless} of the magnitude of the reorganisation energy.

\subsection{Steady state}

We will now look at steady-state accuracy as a function of temperature and system--bath coupling. To do so, we compute the Uhlmann fidelity between the { exact steady state} from Eqs.~\eqref{eq:covariances_coefficients} and the steady state of the weak-coupling master equations discussed above. The fidelity between undisplaced single-mode Gaussian states with covariance matrices $\mathsf{\Sigma}_1$ and $\mathsf{\Sigma}_2$ can be readily calculated using the formula \cite{Scutaru1998fidelity}
\begin{equation*}
	\mathbb F (\mathsf\Sigma_1,\mathsf\Sigma_2) = 2 \left( \sqrt{\varkappa+1}-\sqrt{\Upsilon} \right)^{-1},
\end{equation*}
where $\varkappa = 4 \det{\left(\Sigma_1+\Sigma_2\right)}$ and $\Upsilon =(4 \det{\Sigma_1}-1)(4 \det{\Sigma_2}-1)$. 

As we can see in Fig.~\hyperref[fig2]{2(a)} the Redfield equation derived from $ \pmb{H}_S = \pmb{H}_S^{(0)} + \frac12\delta\omega^2\,\pmb{x}^2 $ gives a good approximation to the mean-force Gibbs state at high-enough temperature and weak-enough coupling $\lambda$. The secular version of such Redfield equation, however, fails miserably, other than at vanishing coupling strengths (cf. Fig.~\hyperref[fig2]{2(b)}). Crucially, as shown in Fig.~\hyperref[fig2]{2(c)}, the long-time limit of the Redfield equation missing both counter term and Lamb-shift contributions provides a very good approximation to $ \pmb{\tau}_{MF} $ almost everywhere in parameter space. Put differently, the high-temperature formula \eqref{eq:semi-classical-state} is remarkably accurate even at low temperatures. 

Our observation that the mean-force Gibbs state can be approximated by the `semi-classical' state in \eqref{eq:semi-classical-state} over a broad range of parameters is of independent interest. Indeed, expressions of the mean-force Gibbs state are only known in the weak and ultra-strong coupling limits \cite{trushechkin2022open,lobejko2022towards,cresser2021weakandultrastrong}, while intermediate couplings may be dealt-with numerically \cite{chiu2022numerical}. Although this may also be model-specific, we remark that the same had been noted for the spin--boson model in Ref.~\cite{timofeev2022hmf}, which raises hopes for a broader applicability. 

\section{Conclusion}\label{sec:conclusions}

We have shown how, whenever the reorganisation energy can be treated perturbatively and the environmental cutoff frequency is sufficiently large, one can derive a second-order Markovian Redfield equation that supports the conventional wisdom within open quantum systems. Namely, the practice of renormalising the Hamiltonian by subtracting the reorganisation energy, and then manually setting the Lamb shift to zero. The resulting master equation accurately tracks the dynamics of any open system irrespective of temperature, so long as the two conditions above are met. The main advantage of this approach is that it singles out the second-order equation asymptotically converging to the classical limit of the correct mean-force Gibbs state. Importantly, such master equation is always accurate at long times, \textit{regardless} of the magnitude of the reorganisation energy, and always provided that one works in the adiabatic regime of large cutoff and the high-temperature limit. This extends to the resulting GKLS obtained after further performing the secular approximation. 

In order to guide the discussion we resorted to the exactly solvable damped quantum harmonic oscillator. In particular, we found that, for this specific system the common practice yields very good results for the transient dynamics and steady state, even when the reorganisation energy is large and the temperature is low. In particular, the agreement between the exact mean-force Gibbs state $\pmb{\tau}_{MF}$ and its classical limit for this particular model, was remarkably good almost everywhere in parmeter space. This extends the regime of applicability the standard \textit{modus operandi} over a broad parameter range for arbitrary networks, made up of linearly coupled harmonic nodes, which encompasses a large class of models of practical interest. 

Two questions remain open. Firstly, we must study the performance of the standard method when evaluating steady-state heat currents for out-of-equilibrium open systems, which are central to quantum thermodynamics. Secondly, it would be interesting to establish whether the classical mean-force Gibbs state remains close to its exact quantum counterpart $\pmb{\tau}_{MF}$ away from the high-temperature limit on other models. Both points certainly deserve further investigation.

\section*{Acknowledgements}

We gratefully acknowledge useful discussions with Janet Anders, Federico Cerisola, James Cresser, Fernando Delgado, Edward Gandar, Charlotte Hogg, José P. Palao, Dvira Segal, Juzar Thingna, Anton Trushechkin, and Alexander White. This work was finished during the \textit{``DQDD Quantum Thermometry Program''} at the University of La Laguna (ULL), funded by ULL and the Spanish Ministry of Universities. LAC is supported by a Ram\'{o}n y Cajal fellowship (RYC2021-325804-I), funded by MCIN/AEI/10.13039/501100011033 and “NextGenerationEU”/PRTR. JG is supported by a scholarship from CEMPS at the University of Exeter.

\appendix

\section{Redfield and GKLS equations}\label{app:redfield}

While we will not elaborate on the well-known steps leading to the Redfield equation (see, e.g., Ref.~\cite{gaspard1999slippage} for a concise and rigorous derivation), we shall provide the explicit equations for the dynamics of the covariances of the system. The equations below have been derived taking $ \pmb{H}_S = \pmb{H}_S^{(0)} $ and inserting the Redfield equation \eqref{eq:Redfield-explicit} into $ \frac{d}{dt}\langle\pmb{O}\rangle = \tr_S(\pmb{O}\,\frac{d}{dt}\pmb{\varrho}_S) $. We thus have
\begin{align*}
	\frac{d}{dt}\langle\pmb{x}\rangle &= \langle\pmb{p}\rangle \\
	\frac{d}{dt}\langle\pmb{p}\rangle &= -(\omega_0^2
	-\Sigma')\,\langle\pmb{x}\rangle + \frac{\Delta}{2\omega_0}\,\langle\pmb{p}\rangle
\end{align*}
for the first-order moments and
\begin{align*}
	\frac{d}{dt}\langle\pmb{x}^2\rangle &= \langle\{\pmb{x},\pmb{p}\}\rangle \\
	\frac{d}{dt}\langle\pmb{p}^2\rangle &= \frac{\Delta}{\omega_0}\,\langle\pmb{p}^2\rangle -(\omega_0^2-\Sigma')\,\langle\{\pmb{x},\pmb{p}\}\rangle + \frac{\Sigma}{2} \\
	\frac{d}{dt}\langle\{\pmb{x},\pmb{p}\}\rangle &= -2(\omega_0^2-\Sigma')\,\langle\pmb{x}^2\rangle + 2\,\langle\pmb{p}^2\rangle +\frac{\Delta}{2\omega_0}\,\langle\{\pmb{x},\pmb{p}\}\rangle -\frac{\Delta'}{\omega_0}
\end{align*}
for the second-order moments.

Using instead the secular GKLS equation \eqref{eq:Lindblad} gives
\begin{align*}
	\frac{d}{dt}\langle\pmb{x}\rangle &=  \frac{\Delta}{4\omega_0}\langle\pmb{x}\rangle + \frac{1}{\omega_0^2}\,\left(\omega_0^2-\frac{\Sigma'}{2}\right)\,\langle\pmb{p}\rangle\\
	\frac{d}{dt}\langle\pmb{p}\rangle &=  -\left(\omega_0^2 -\frac{\Sigma'}{2}\right)\,\langle\pmb{x}\rangle + \frac{\Delta}{4\omega_0}\,\langle\pmb{p}\rangle \\
	\frac{d}{dt}\langle\pmb{x}^2\rangle &= \frac{\Delta}{2\omega_0}\,\langle\pmb{x}^2\rangle + \frac{1}{\omega_0^2}\,\left(\omega_0^2-\frac{\Sigma'}{2}\right)\,\langle\{ \pmb{x},\pmb{p} \}\rangle - \frac{\Sigma}{4\omega_0^2}\\
	\frac{d}{dt}\langle\pmb{p}^2\rangle &= \frac{\Delta}{2\omega_0}\,\langle\pmb{p}^2\rangle - \left(\omega_0^2-\frac{\Sigma'}{2}\right)\,\langle\{ \pmb{x},\pmb{p} \}\rangle - \frac{\Sigma}{4} \\
	\frac{d}{dt}\langle\{\pmb{x},\pmb{p}\}\rangle &= -(2\omega_0^2-\Sigma')\,\langle\pmb{x}^2\rangle + \frac{2}{\omega_0^2}\,\left(\omega_0^2-\frac{\Sigma'}{2}\right)\,\langle\pmb{p}^2\rangle \\&\qquad\qquad\qquad\qquad\qquad\qquad\quad + \frac{\Delta}{2\omega_0}\,\langle\{\pmb{x},\pmb{p}\}\rangle.
\end{align*}

\section{Detailed solution of the exact dynamics}
\label{app:Heisenberg}

The Heisenberg equations of motion from the Hamiltonian $\pmb{H}_S + \pmb{V} + \pmb{H}_B $ defined in Eqs.~\eqref{eq:system_Hamiltonian} and \eqref{eq:sys-bath} are
\begin{align*}
	\dot{\pmb x} &= \pmb p,\\
	\dot{\pmb p} &= - \omega_0^2\,\pmb x - \sum\nolimits_\mu g_\mu\,\pmb x_\mu,\\
	\dot{\pmb x}_\mu &= \pmb p_\mu/m_\mu,\\
	\dot{\pmb p}_\mu &= - m_\mu \omega_\mu^2\,\pmb x_\mu - g_\mu\,\pmb x.
\end{align*}    
In order to simplify notation we define $ \pmb{\mathsf{Z}}_\mu = (\pmb x_\mu, \pmb p_\mu)^\mathsf{T}$ and $ \pmb{\mathsf{Z}} = (\pmb x, \pmb p)^\mathsf{T} $. Hence, the equation of motion for the environmental degrees of freedom can be compactly expressed as
\begin{equation*}
	\dot{\pmb{\mathsf{Z}}}_\mu(t) + \mathsf{\Omega}_\mu \pmb{\mathsf{Z}}_\mu(t)  = \mathsf{C}_\mu \pmb{\mathsf{Z}}(t),
\end{equation*}
with 
\begin{align*}
	\mathsf{\Omega}_\mu &= \begin{pmatrix}0 & - m_\mu^{-1}\\ m_\mu \omega_\mu^2&0\end{pmatrix} \\
	\mathsf{C}_\mu &= \begin{pmatrix}0 & 0\\ -g_\mu&0\end{pmatrix}.   
\end{align*}

Here $\pmb{\mathsf{Z}}(t)$ acts as a source term in the equations for $\pmb{\mathsf{Z}}_\mu(t)$. Taking the Laplace transform of Eq.~\eqref{eq:vectorised-eom-probe} we obtain
\begin{equation*}
	s\,\widehat{{\pmb{\mathsf{Z}}}}_\mu(s)- \pmb{\mathsf{Z}}_\mu(0) + \mathsf{\Omega}_\mu \widehat{{\pmb{\mathsf{Z}}}}_\mu(s)= \mathsf{C}_\mu \widehat{{\pmb{\mathsf{Z}}}}(s) ,
\end{equation*}
which can be recast as
\begin{equation*}
	\widehat{{\pmb{\mathsf{Z}}}}_\mu(s)= \left(s\,\mathbbm{1}_2 + \mathsf{\Omega}_\mu \right)^{-1} \left[\mathsf{C}_\mu \widehat{{\pmb{\mathsf{Z}}}}(s) + \pmb{\mathsf{Z}}_\mu(0)\right],
\end{equation*}
where $\mathbbm{1}_2$ is the $2\times 2$ identity matrix. Taking now the inverse Laplace transform we get
\begin{equation*}
	\pmb{\mathsf{Z}}_\mu(t) = \mathsf{G}_\mu(t)\, \pmb{\mathsf{Z}}_\mu(0) + \int_0^t dt'\,\mathsf{G}_\mu(t-t')\,\mathsf{C}_\mu\,\pmb{\mathsf{Z}}(t').
\end{equation*}
Here, the elements of $ \mathsf{G}_\mu(t) $ are 
\begin{equation*}
	\mathsf{G}_\mu(t)  = \begin{pmatrix}\cos{(\omega_\mu t)} & \frac{\sin{(\omega_\mu t)}}{m_\mu\,\omega_\mu} \\-m_\mu \omega_\mu \sin{(\omega_\mu t)}&\cos{(\omega_\mu t)}\end{pmatrix}.
\end{equation*}

Using this, we can now express the equations of motion of the degrees of freedom of the system as
\begin{equation}
	\dot{\pmb{\mathsf{Z}}}(t) + \mathsf{\Omega}\,\pmb{\mathsf{Z}}(t) - \int_0^t dt'\, \mathsf{X}(t-t')\,\pmb{\mathsf{Z}}(t')  = \pmb{\mathsf{F}}(t) \label{eq:vectorised-eom-probe},
\end{equation}
with
\begin{align*}
	\mathsf{\Omega} &= \begin{pmatrix}0 & -1\\ \omega_0^2 &0\end{pmatrix} \\
	\mathsf{X}(t)&=\begin{pmatrix}0 & 0\\\chi(t)&0\end{pmatrix},
\end{align*}
where the noise term is $\pmb{\mathsf{F}}(t) = \sum_\mu \mathsf{C}_\mu\,\mathsf{G}_\mu(t)\,\pmb{\mathsf{Z}}_\mu(0)$. Taking again the Laplace transform of \eqref{eq:vectorised-eom-probe} we obtain
\begin{equation*}
	s\,\widehat{{\pmb{\mathsf{Z}}}}(s)- \pmb{\mathsf{Z}}(0) + \mathsf{\Omega}\,\widehat{{\pmb{\mathsf{Z}}}}(s) -  \widehat{\mathsf{X}}(s)\,\widehat{{\pmb{\mathsf{Z}}}}(s)  = \widehat{\pmb{\mathsf{F}}}(s),
\end{equation*}
which can be recast as
\begin{equation*}
	\widehat{{\pmb{\mathsf{Z}}}}(s)= \left(s\,\mathbbm{1}_2 + \mathsf{\Omega}  -  \widehat{\mathsf{X}}(s)\right)^{-1} \left(\widehat{\pmb{\mathsf{F}}}(s) + \pmb{\mathsf{Z}}(0)\right).
\end{equation*}
Transforming this back into the time domain takes us to Eqs.~\eqref{eq:G-t} from the main text. In turn, these are all we need to calculate the correlation functions of our system.

In particular, the covariance matrix \eqref{eq:covariance-matrix-general} evolves as
\begin{align}\label{eq:covariance-time}
	\mathsf{\Sigma}(t) = \mathsf{G}(t)\,\mathsf{\Sigma}(0)\,\mathsf{G}(t)^\mathsf{T} + \mathsf{G}(t)\,\textrm{Re}\,\langle \pmb{\mathsf{Z}}(0)\pmb{\mathsf{B}}(t)^\mathsf{T} \rangle
	+ \textrm{Re}\,\langle \pmb{\mathsf{B}}(t)\pmb{\mathsf{Z}}(0)^\mathsf{T} \rangle \,\mathsf{G}(t)^\mathsf{T} + \textrm{Re}\,\langle\pmb{\mathsf{B}}(t)\pmb{\mathsf{B}}(t)^\mathsf{T} \rangle,
\end{align}
with $\pmb{\mathsf{B}}(t)= \int_0^\infty\,\mathsf{G}(t-t')\,\pmb{\mathsf{F}}(t')\,dt'$. The second and third term are propagations of the initial system--bath correlations and therefore vanish for our factorised preparation. In turn, the first term vanishes asymptotically, so that it must be considered when studying the dynamics of the system. We now turn our attention to the more complicated fourth term, which we dub `memory integral' term. Explicitly, this is given by
	\begin{multline}\label{eq:limit-cycle-covariance-raw}
		\mathsf{\Sigma}_\text{mi}(t) = \textrm{Re}\, \langle \pmb{\mathsf{B}}(t)\pmb{\mathsf{B}}(t)^\mathsf{T}\rangle \\= \int_0^t dt_1 \int_0^t dt_2\,\mathsf{G}(t-t_1) \sum\nolimits_{\mu\nu} \mathsf{C}_\mu \mathsf{G}_\mu(t_1)\,\textrm{Re}\, \langle \pmb{\mathsf{Z}}_\mu(0)\pmb{\mathsf{Z}}_\nu(0)^\mathsf{T} \rangle \mathsf{G}_\nu(t_2)^\mathsf{T} \mathsf{C}^\mathsf{T}_\nu \mathsf{G}(t-t_2)^\mathsf{T}\, .
	\end{multline}
Since the bath is in thermal equilibrium at temperature $T$
\begin{align*}
	\textrm{Re}\,\langle \pmb{\mathsf{Z}}_\mu(0)\,\pmb{\mathsf{Z}}_\nu(0)^\mathsf{T} \rangle = \delta_{\mu\nu} \begin{pmatrix} \frac{1}{2\,m_\mu \omega_\mu} \coth\frac{\omega_\mu}{2 T}&0\\0 &\frac{m_\mu \omega_\mu}{2} \coth\frac{\omega_\mu}{2 T}\end{pmatrix}, 
\end{align*}
where $\delta_{\mu\nu}$ is a Kronecker delta. Eq.~\eqref{eq:limit-cycle-covariance-raw} thus simplifies to
\begin{equation}\label{eq:covariance-limit-cycle_general}
	\mathsf{\Sigma}^\text{mi}(t) = \frac12\,\int_0^t dt_1 \int_0^t dt_2\,\mathsf{G}(t-t_1)\begin{pmatrix} 0&0\\0 &\mu(t_1-t_2)\end{pmatrix} \mathsf{G}(t-t_2)^\mathsf{T},
\end{equation}
with the ‘noise kernel’ $ \mu(t) $ given by
\begin{align*}
	\mu(t) = \sum\nolimits_\mu \frac{g_\mu^2}{m_\mu\omega_\mu}\,\cos{\omega_\mu t}\,\coth{\frac{\omega_\mu}{2 T}}
	= \frac{2}{\pi}\,\int_0^\infty d\omega\,J(\omega)\,\cos{\omega t}\, \coth{\frac{\omega}{2 T}}.
\end{align*}

Finally, Eq.~\eqref{eq:covariance-limit-cycle_general} reduces to the following expressions for the memory-integral contribution to the elements of $\mathsf{\Sigma}(t)$:
\begin{align*}
	[\mathsf{\Sigma}^{\text{mi}}(t)]_{11} &= \frac{1}{2}  \int_0^t dt_1 \int_0^t dt_2~g(t-t_1)\,\mu(t_1-t_2)\,g(t-t_2), \\
	[\mathsf{\Sigma}^{\text{mi}}(t)]_{22}  &= \frac{1}{2}  \int_0^t dt_1 \int_0^t dt_2~\partial_t\,g(t-t_1)\,\mu(t_1-t_2)\,\partial_t g(t-t_2),\\
	[\mathsf{\Sigma}^{\text{mi}}(t)]_{12} &= \frac{1}{2}  \int_0^t dt_1 \int_0^t dt_2~g(t-t_1)\,\mu(t_1-t_2)\,\partial_t g(t-t_2).
\end{align*}
Note that since the inverse Laplace transform is given by the Bromwich integral, the diagonal elements of $\mathsf{G}(t)$ can be cast as $ [\mathsf{G}(t)]_{11} = [\mathsf{G}(t)]_{22} = \partial_t g(t)$.

In order to study the steady state, it is convenient to write the covariances as
\begin{equation*}
	[\mathsf{\Sigma}^{\text{mi}}(t)]_{11} = \langle \pmb{x}^2 (t)\rangle^\text{mi} = \int_0^\infty d\omega\,\nu(\omega)\,\varsigma_{xx}(t,\omega)
\end{equation*}
where $\nu(\omega)$ was defined in Eq.~\eqref{eq:nk} and thus,
\begin{align*}
	\varsigma_{xx}(t,\omega) &= \frac{1}{2} \int_0^t dt_1 \int_0^t dt_2~g(t-t_1)\,\cos{[\omega(t_1-t_2)]}\,g(t-t_2) \\
	&= \frac{1}{4}  \int_0^t  du_1~g(u_1)\,e^{- i\,\omega\,u_1 } \int_0^t du_2~ g(u_2)\,e^{ i\,\omega\,u_2 } + \text{c.c.}.
\end{align*}
Here, we have used Euler's formula, $\cos{x}=\frac{1}{2} (e^{i\,x} + e^{-i\,x})$, and introduced the variables $u_j=t-t_j$. Asymptotically, this gives
\begin{align*}
	\lim\nolimits_{t \rightarrow \infty}\,\varsigma_{xx}(t,\omega) &= \frac12~\widehat{g}(-i\,\omega)~\widehat{g}(i\,\omega), 
\end{align*}
which brings us to Eq.~\eqref{eq:xx_ss} in the main text; namely, 
\begin{equation*}
	\langle \pmb{x}^2 \rangle_\infty =\frac{1}{2}\int_0^\infty  \vert \widehat{g}(i \omega )\vert^2\,\nu(\omega)\,d\omega.
\end{equation*}
Note that we have set the first term in Eq.~\eqref{eq:covariance-time} to zero (i.e., $ \langle \pmb{x}^2 \rangle^\text{mi}_\infty = \langle \pmb{x}^2 \rangle_\infty $), since it vanishes in steady state.

Now, using the fact that $g(0) = [\mathsf{G}(0)]_{12} = 0$, we see that $ \widehat{\partial_t\,g}(s) = s\,\widehat{g}(s) $, which gives a vanishing asymptotic position--momentum covariance $ \langle \{\pmb{x},\pmb{p}\} \rangle_\infty = 0 $, since
\begin{align*}
	\lim\nolimits_{t \rightarrow \infty} \varsigma_{xp}(t,\omega) = \frac{1}{4} \big[i\,\omega\,\vert\widehat{g}(i\,\omega)\vert^2- i\,\omega\,\vert\widehat{g}(i\,\omega)\vert^2\big] = 0.
\end{align*}
Finally, for momentum variance we have
\begin{align*}
	\lim\nolimits_{t \rightarrow \infty} \varsigma_{pp}(t,\omega) = \frac12\,\omega^2\,\vert\widehat{g}(i\,\omega)\vert^2,
\end{align*}
so that we recover Eq.~\eqref{eq:pp_ss}.

\renewcommand{\thefootnote}{\fnsymbol{footnote}}

\bibliography{references}

\end{document}